\documentclass[11pt]{article}
\usepackage{graphicx}
\usepackage{amssymb}
\usepackage{amscd}
\usepackage{mathrsfs}
\usepackage{longtable,lscape}
\usepackage{amsthm}
\usepackage{amsfonts}
\usepackage{amsmath}
\usepackage{bbm}
\usepackage{float}
\usepackage{url}

\newcommand{\compl}{{\mathbb C}}
\newcommand{\real}{{\mathbb R}}
\newcommand{\captionfonts}{\footnotesize}
\makeatletter  
\long\def\@makecaption#1#2{%
  \vskip\abovecaptionskip
  \sbox\@tempboxa{{\captionfonts #1: #2}}%
  \ifdim \wd\@tempboxa >\hsize
    {\captionfonts #1: #2\par}
  \else
    \hbox to\hsize{\hfil\box\@tempboxa\hfil}%
  \fi
  \vskip\belowcaptionskip}
\makeatother 
\begin{document}
\title{The Extended Bloch Representation of Quantum Mechanics for Infinite-Dimensional Entities}
\author{Diederik Aerts and Massimiliano Sassoli de Bianchi \vspace{0.5 cm} \\ 
        \normalsize\itshape
        Center Leo Apostel for Interdisciplinary Studies, 
         Brussels Free University \\ 
        \normalsize\itshape
         Krijgskundestraat 33, 1160 Brussels, Belgium \\
        \normalsize
        E-Mails: \url{diraerts@vub.ac.be,msassoli@vub.ac.be}
          \vspace{0.5 cm} \\ 
              }
\date{}
\maketitle
\begin{abstract}
\noindent
We show that the extended Bloch representation of quantum mechanics also applies to infinite-dimensional entities, to the extent that the number of (possibly infinitely degenerate) outcomes of a measurement remains finite, which is always the case in practical situations. 
\end{abstract}
\maketitle

\section{Introduction}
\label{Introduction}
The so-called `spin quantum machine', also known as the `$\epsilon$-model', or `sphere model' \cite{AertsEtAl1997,Aerts1998,Aerts1999}, is an extension of the standard ($3$-dimensional) Bloch sphere representation that includes a description also of the measurements, as (weighted) symmetry breaking processes selecting (in a non-predictable way) the hidden-measurement interactions responsible for producing the transitions towards the outcome-states. Recently, the model has been extended, so that measurements having an arbitrary number $N$ of (possibly degenerate) outcomes can also be described, in what has been called the `extended Bloch representation' (EBR) of quantum mechanics \cite{asdb2014,asdb2015a,asdb2015b,asdb2016a,asdb2016b,asdb2016d,asdb2017}.

So far, the EBR has been formulated only for finite-dimensional quantum entities, although of arbitrary dimension. It is thus natural to ask if the representation remains consistent when dealing with infinite-dimensional entities. Of course, certain quantum entities, like spin entities, are intrinsically finite-dimensional. For instance, the Hilbert state space of a spin-$s$ entity is $(2s+1)$-dimensional and can be taken to be isomorphic to ${\cal H}=\compl^{2s+1}$. However, an entity as simple as an electron, when considered in relation to position or momentum measurements, requires an infinite-dimensional Hilbert space ${\cal H}=L^2(\real^3)$, to account for all its possible states.

As we emphasized in \cite{asdb2014}, the EBR being valid for an arbitrary finite number $N$ of dimensions, it can be advocated that if the physics of infinite-dimensional entities can be recovered by taking the limit $N\to\infty$ of suitably defined finite-dimensional entities, then a hidden-measurement description of quantum measurements should also apply for infinite-dimensional entities. More precisely, assuming that it is always possible to express the transition probabilities of an infinite-dimensional entity as the limit of a sequence of transition probabilities of finite-dimensional entities, and considering that a hidden-measurement interpretation holds for the latter, one would expect it to also hold for the former. 

However, the possibility of a hidden-measurement interpretation does not necessarily imply the existence, for infinite-dimensional entities, of an explicit representation. For instance, some years ago Coecke was able to apply the hidden-measurement approach to measurements having an infinite set of outcomes, but to do so he had to take the space describing the hidden-measurement interactions to be the fixed interval $[0,1]$, independently of the number $N$ of outcomes, and this precisely to avoid problems when taking the infinite limit $N\to\infty$ \cite{Coecke1995b}. His construction is thus very different from the canonical EBR, as in the latter the dimension of the set of hidden-measurements depends on the number of outcomes. 

More precisely, in the EBR the set of hidden-measurement interactions, for a measurement having $N$ possible outcomes, is given by a $(N-1)$-dimensional simplex $\triangle_{N-1}$, inscribed in a convex region of states which, in turn, is inscribed in a $(N^2-1)$-dimensional unit sphere $B_1(\real^{N^2-1})$ \cite{asdb2014}. A measurement then consists first in a deterministic process, producing a decoherence of the pre-measurement state, represented by an abstract point at the surface of the sphere that plunges into it, to reach the measurement simplex $\triangle_{N-1}$, following a path orthogonal to the latter. In this way, $N$ different disjoint subregions $A_i$ of $\triangle_{N-1}$ are defined, $i=1,\dots,N$, whose measures\footnote{One should say, more precisely, $(N-1)$-dimensional volumes, or Lebesgue measures.} describe the number of measurement-interactions that are available to actualize the corresponding outcomes. This means that the relative measures ${\mu(A_i)\over\mu(\triangle_{N-1})}$ of the different subregions can be interpreted as the probabilities to obtain the associated outcomes, and the remarkable result of the model is that these probabilities are exactly those predicted by the Born rule \cite{asdb2014}. 

However, when taking the $N\to\infty$ limit of the extended Bloch construction, there is the following problem. The $M$-dimensional volume of a $M$-ball of radius $r$, given by:
\begin{eqnarray}
\mu[B_r(\real^{M})]={\pi^{M\over 2}r^M\over \Gamma ({\frac{M}{2}}+1)},
\label{volume-sphere}
\end{eqnarray}
tends to zero, as $M\to\infty$. Indeed, if $M$ is even, we have $\Gamma ({\frac{M}{2}}+1)={M\over 2}!$, so that according to Stirling's approximation, $M!\sim \sqrt{2\pi M}({M\over e})^M$, we have the asymptotic behavior:
\begin{eqnarray}
\mu[B_r(\real^{M})] \sim {1\over \sqrt{2e} \pi r}\left({\sqrt{2\pi e \over M}\, r}\right)^{M+1},
\label{volume-sphere-asymptotic}
\end{eqnarray}
as $M\to\infty$, and a similar asymptotic formula can be found when $M$ is odd, using $\Gamma ({\frac{M}{2}}+1)=\sqrt{\pi}\, 2^{-{M+1\over 2}}M!!$. In other words, the measure of a $M$-dimensional ball of fixed radius $r$ goes to zero extremely fast when the dimension $M$ increases.\footnote{The fact the measure of a $M$-ball (its $M$-dimensional volume) of fixed radius tends exponentially fast to zero as $M$ increases is counter intuitive. Indeed, for a unit radius $r=1$, we have $\mu[B_1(\real^{1})] =2$, $\mu[B_1(\real^{2})] =\pi >2$, $\mu[B_1(\real^{3})] ={4\over 3}\pi >\pi$, $\mu[B_1(\real^{4})] ={\pi^2\over 2}>{4\over 3}\pi$, $\mu[B_1(\real^{5})] ={8\pi^2\over 15}>{\pi^2\over 2}$, but $\mu[B_1(\real^{6})] ={\pi^3\over 6}<{8\pi^2\over 15}$. In other words, the measure increases from $M=1$ to $M=5$, then it starts decreasing as from $M=6$.} 

The same is necessarily true for all structures of same dimension that are contained in it, like for instance inscribed simplexes. More precisely, the measure of a $M$-dimensional simplex $\triangle_{M}$, inscribed in a sphere of radius $r$ is \cite{asdb2014}: 
\begin{equation}
\mu(\triangle_{M})={\sqrt{M}\over M!}\left({M+1\over M}\right)^{M+1\over 2}r^{M}.
\end{equation}
Using again Stirling's approximation, we thus obtain the asymptotic form: 
\begin{equation}
\mu(\triangle_{M})\sim {1\over \sqrt{2\pi}}\left({e\, r\over M}\right)^{M},
\end{equation}
which goes even faster to zero than (\ref{volume-sphere-asymptotic}). So, if we naively consider the infinite-dimensional limit of the EBR, we find that the measures of the structures involved in the model rapidly go to zero. Nevertheless, considering that in actual measurement situations the number of distinguishable outcomes is always finite, this will prove to be unproblematic, as we are going to show. Also, as we will suggest in the last section, one can even speculate that the measurement-interactions would precisely supervene because of the meeting between an entity that is possibly infinite-dimensional, and the constraints exercised by a measurement context only allowing for a finite number of possible outcomes.

\section{The infinite-dimensional limit}

According to the EBR of quantum mechanics, the transition probability ${\cal P}[D_N({\bf r})\to P_N({\bf n})]$, from an initial state $D_N({\bf r})$ to a final outcome-state $P_N({\bf n}_i)$, is given by the formula \cite{asdb2014}: 
\begin{eqnarray}
{\cal P}[D_N({\bf r})\to P_N({\bf n}_i)] = {1\over N} \left[1+ (N-1)\,{\bf r}\cdot {\bf n}_i\right]={1\over N} \left[1+ (N-1)\,{\bf r}^\parallel\cdot {\bf n}_i\right].
\label{transitiongeneralNxN}
\end{eqnarray}
Here $D_N({\bf r})$ and $P_N({\bf n}_i)$ are one-dimensional projection operators acting in ${\cal H}_N=\compl^N$, which can be written as:
\begin{eqnarray}
D_N({\bf r})={1\over N}(\mathbb{I}_N + c_N\,{\bf r}\cdot \mbox{\boldmath$\Lambda$}),\quad P_N({\bf n}_i)={1\over N}(\mathbb{I} + c_N\,{\bf n}_i\cdot \mbox{\boldmath$\Lambda$}),
\label{exp}
\end{eqnarray}
where ${\bf r}$ and ${\bf n}_i$ are unit vectors in the generalized Bloch sphere $B_1(\real^{N^2-1})$, with ${\bf n}_i$ being also one of the $N$ vertices of a given $(N-1)$-dimensional measurement simplex $\triangle_{N-1}$, inscribed in $B_1(\real^{N^2-1})$, $c_N= [N(N-1)/2]^{1\over 2}$, \mbox{\boldmath$\Lambda$} is a vector whose components $\Lambda_i$ are a choice of the $N^2-1$ generators of the group $SU(N)$, and the `dot' denotes the scalar product in $\real^{N^2-1}$. In (\ref{transitiongeneralNxN}) we have also introduced the vector ${\bf r}^\parallel={\bf r}-{\bf r}^\perp$, where ${\bf r}^\perp$ is the component of ${\bf r}$ perpendicular to $\triangle_{N-1}$, i.e., ${\bf r}^\perp\cdot {\bf n}_i=0$, for all $i=1,\dots,N$ (we refer the reader to \cite{asdb2014} for a detailed exposition).

It is straightforward to take the $N\to\infty$ limit of (\ref{transitiongeneralNxN}). By doing so, one just needs to keep in mind that also the dimension of the Hilbert space increases, as $N$ increases. In other words, one has to assume that, as $N\to\infty$, both $D_N({\bf r})$ and $P_N({\bf n}_i)$ converge (in the Hilbert-Schmidt sense) to well-defined projection operators $D({\bf r})$ and $P({\bf n}_i)$, respectively, acting in ${\cal H}_\infty = \ell^2(\compl)$.\footnote{$\ell^2(\compl)$ is the Hilbert space of infinite sequences $\{v_0,v_1,\dots\}$ of complex numbers satisfying $\sum_{i=0}^{\infty}|v_i|^2<\infty$, with scalar product $\langle v| w\rangle = \sum_{i=0}^{\infty}v_i^*w_i$.} More precisely, we have to assume that: 
\begin{eqnarray}
\left|{\cal P}(D_N\to P_N)-{\cal P}(D\to P)\right| &=& \left| {\rm Tr}\, D_N P_N -{\rm Tr}\, D P\right| = \left|{\rm Tr}\, D_N (P_N-P) + {\rm Tr}\, (D_N -D)P \right|\nonumber \\
&\leq& \left|{\rm Tr}\, (P_N-P)\right| + \left|{\rm Tr}\, (D_N -D) \right|\to 0,\,\, {\rm as}\,N\to\infty.
\end{eqnarray}
Then, the $N\to\infty$ limit of (\ref{transitiongeneralNxN}) is: 
\begin{equation}
{\cal P}[D({\bf r})\to P({\bf n}_i)] = {\bf r}\cdot {\bf n}_i = {\bf r}^\parallel \cdot {\bf n}_i,
\label{transitioninfty}
\end{equation}
where ${\bf r}$ and ${\bf n}_i$ are now vectors belonging to $\ell^2(\real)$, the Hilbert space of infinite sequences $\{r_1,r_1,\dots\}$ of real numbers satisfying $\sum_{i=1}^{\infty}r_i^2<\infty$, with scalar product ${\bf r}\cdot {\bf n}_i = \sum_{j=1}^{\infty}r_j [{\bf n}_i]_j$.

It is worth remembering that one of the differences between the standard Bloch representation ($N=2$) and the EBR, for $N>2$, is that in the latter not all vectors in $B_{1}(\real^{N^2-1})$ are necessarily representative of \emph{bona fide} states. However, all good states are represented by vectors belonging to a convex region inscribed in $B_{1}(\real^{N^2-1})$. The vectors living outside of such convex region of states (the shape of which depends on the choice of the generators $\Lambda_i$) can be characterized by the fact that for them (\ref{transitiongeneralNxN}) would give unphysical negative values. This possibility is even more manifest in the infinite-dimensional formula (\ref{transitioninfty}), as is clear that a scalar product can take both positive and negative values. For instance, the unit vector ${\bf r}=- {\bf n}_i$ cannot be representative of a state, as for it the transition probability (\ref{transitioninfty}) would be equal to $-1$.

Let us now investigate what is the $N\to\infty$ limit of the $(N-1)$-dimensional measurement simplex $\triangle_{N-1}$. By definition, we have:
\begin{equation}
\triangle_{N-1}=\{{\bf t}\in\real^N | {\bf t}=\sum_{i=1}^{N} t_i\, {\bf n}_i,\, \sum_{i=1}^N t_i = 1, 0\leq t_i\leq 1\},
\label{simplex}
\end{equation}
where the ${\bf n}_i$, $i=1,\dots,N$, are the $N$ vertex vectors of $\triangle_{N-1}$, describing the $N$ outcome states and obeying:
\begin{equation}
{\bf n}_i\cdot{\bf n}_j=-{1\over N-1}+\delta_{ij}{N\over N-1},
\label{scalar}
\end{equation}
so that we also have $\sum_{i=1}^N {\bf n}_i = {\bf 0}$. Taking the $N\to\infty$ limit of (\ref{simplex})-(\ref{scalar}), we thus obtain: 
\begin{equation}
\triangle_{\infty}=\{{\bf t}\in\real^\infty | {\bf t}=\sum_{i=1}^\infty t_i\, {\bf n}_i,\, \sum_{i=1}^\infty t_i = 1, 0\leq t_i\leq 1\},
\label{simplex-infinite}
\end{equation}
with the outcome states now obeying:
\begin{equation}
{\bf n}_i\cdot{\bf n}_j=\delta_{ij},
\label{scalar-infinite}
\end{equation}
i.e., they are all mutually orthogonal unit vectors in $\ell^2(\real)$.

Clearly, ${\bf 0}\in \triangle_{N-1}$, for all $N<\infty$, i.e., finite-dimensional simplexes contain the null vector ${\bf 0}$, which describes their center, representative of the operator-state ${1\over N}\mathbb{I}_N$. For instance, for the $N=2$ case, taking $s_1=s_2={1\over 2}$, and considering that ${\bf n}_1=-{\bf n}_2$, we clearly have ${\bf 0}={1\over 2} {\bf n}_1+ {1\over 2}{\bf n}_2\in \triangle_{1}$. On the other hand, vectors belonging to $\triangle_{\infty}$ are convex combinations of mutually orthogonal unit vectors, so that ${\bf 0}\notin \triangle_{\infty}$. In other words, by taking the infinite limit we shift from a representation where the null vector is the center of the simplexes, to a standard (infinite) representation where the null vector describes a point external to the simplex (see Appendix~\ref{Appendix}). 

So, given a (pure point spectrum) observable $A$, acting in ${\cal H}_\infty = \ell^2(\compl)$, with spectral family $\{P({\bf n}_1),P({\bf n}_2),\dots\}$, where the $P({\bf n}_i)$ are mutually orthogonal one-dimensional projection operators, ${\rm Tr}\, P({\bf n}_i)P({\bf n}_j) = \delta_{ij}\,\mathbb{I}$, we can associate to it an infinite dimensional (standard) simplex $\triangle_{\infty}$, with vertices ${\bf n}_i\cdot {\bf n}_j=\delta_{ij}$, in such a way that the transition probabilities from an operator-state $D({\bf r})$ to the vector-eigenstates $P({\bf n}_i)$ are simply given by the (real) scalar products (\ref{transitioninfty}) [see also (\ref{transitiongeneralNxN2-bis})]. Also, to each vector ${\bf n}_i$, we can associate a region $A_i\subset \triangle_{\infty}$, corresponding to the convex closure of $\{{\bf n}_1, \dots, {\bf n}_{i-1}, {\bf r}^\parallel, {\bf n}_{i+1}, \dots\}$. However, since $\mu(A_{i})= \mu(\triangle_{\infty})=0$, we cannot anymore define the transition probabilities ${\cal P}[D({\bf r})\to P({\bf n}_i)]$ as the ratios ${\mu(A_{i})\over \mu(\triangle_{\infty})}$, as they are now undefined ``zero over zero'' ratios.
So, different from the finite-dimensional situation, it seems not to be anymore possible to understand the scalar product (\ref{transitioninfty}) as resulting from the processes of actualization of the available potential measurement-interactions. 

Of course, as we mentioned already in Sec.~\ref{Introduction}, it is always possible to understand ${\mu(A_{i})\over \mu(\triangle_{\infty})}$ as the limit of well-defined ratios, associated with finite-dimensional systems, considering that the EBR works for all finite $N$. However, we would like to elucidate if a hidden-measurement mechanism can also be directly highlighted for infinite-dimensional entities. More precisely, can we maintain that, when the abstract point particle representative of the state, initially located in ${\bf r}$, orthogonally ``falls" onto the measurement simplex $\triangle_{\infty}$, thus producing the deterministic (decoherence-like) transition ${\bf r}\to {\bf r}^\parallel$, a subsequent indeterministic process takes place, describable as a weighted symmetry breaking over the available hidden-measurement interactions, in accordance with the Born rule?

To answer this question, we start by considering the simple situation where ${\bf r}^\parallel$ can be written as the convex combination of only two vertex vectors, i.e., ${\bf r}^\parallel=r_i^\parallel\, {\bf n}_i + r_j^\parallel\, {\bf n}_j$, for some $i$ and $j$, $i\neq j$. In other words, we assume that following the transition ${\bf r}\to {\bf r}^\parallel$, the on-simplex vector ${\bf r}^\parallel$ belongs to one of the edges of the infinite simplex, i.e., ${\bf r}^\parallel\in \tilde{\Delta}_{1}({\bf n}_i, {\bf n}_j)=\{{\bf t}\in\real^2 | {\bf t}=t_i \, {\bf n}_i + t_j\, {\bf n}_j,\, t_i + t_j = 1, 0\leq t_i\leq 1\}$, with ${\bf n}_i\cdot{\bf n}_j=0$.\footnote{We have introduced the notation $\tilde{\Delta}$ (with a tilde) to indicate that, contrary to (\ref{simplex}), the simplex is a standard one, defined in terms of orthonormal vertex vectors.} The measure of $\tilde{\Delta}_{1}({\bf n}_i, {\bf n}_j)$ is of course finite and equal to the edge's length, i.e., $\mu[\tilde{\Delta}_{1}({\bf n}_i, {\bf n}_j)]=\|{\bf n}_j - {\bf n}_i\|=\sqrt{2}$, and can be considered to be representative of the available measurement-interactions. Also, we have that $A_i$ is the convex closure of $\{{\bf r}^\parallel, {\bf n}_j\}$, and $A_j$ is the convex closure of $\{{\bf r}^\parallel, {\bf n}_i\}$, so that:
\begin{eqnarray}
\mu(A_j)= \|{\bf r}^\parallel-{\bf n}_i\|=\|(r_i^\parallel-1)\,{\bf n}_i + r_j^\parallel\,{\bf n}_j\|= \|r_j^\parallel \, ({\bf n}_j - {\bf n}_i)\| = r_j^\parallel \sqrt{2},
\end{eqnarray}
and similarly: $\mu(A_i)= r_i^\parallel \sqrt{2}$. Thus, in accordance with the Born rule, we have the well-defined ratio: 
\begin{eqnarray}
{\cal P}[D({\bf r})\to P({\bf n}_k)] = {\mu(A_k)\over \mu[\tilde{\Delta}_{1}({\bf n}_i, {\bf n}_j)]}= {\|{\bf r}^\parallel-{\bf n}_k\|\over \|{\bf n}_j - {\bf n}_i\|} = {r_k^\parallel \sqrt{2}\over \sqrt{2}} = r_k^\parallel,\quad k=i,j.
\end{eqnarray}

The above simple exercise was to emphasize that the logic of the EBR remains intact also when working with the infinite (standard) simplex $\Delta_\infty$, if the initial state vector is the convex combination ${\bf r}^\parallel=r_i^\parallel\, {\bf n}_i + r_j^\parallel\, {\bf n}_j$, and of course the same reasoning also applies, \emph{mutatis mutandis}, when ${\bf r}^\parallel$ is the convex combination of a finite arbitrary number $N$ of vertex vectors. However, these are very special circumstances, as in general ${\bf r}^\parallel$ will be written as a convex combination of an infinite number of vertex vectors, and in that case we face the previously mentioned difficulty that $\mu(\Delta_\infty)=0$. But, is it really so? 

It is worth making the distinction between the fact that an infinite-dimensional quantum entity can in principle produce an infinity of outcome-states and the fact that in actual experimental situations not all these \emph{a priori} possible outcome-states will be truly available to be actualized. In other words, actual measurement contexts, like those we create in our laboratories, only allow for a finite number of possible outcomes, because the number of detectors, however large, is necessarily finite, and their resolving power is also limited. This means that, even though the dimension of a quantum entity, like an electron, can be infinite, its measurement contexts are always finite-dimensional, and therefore described by degenerate measurements. This means that actual measurements need to be associated with effective finite-dimensional simplexes, for which the hidden-measurement interpretation always applies in a consistent way. Let us show how this works. 

We introduce $N$ different disjoint subsets $I_{i}$ of $\mathbb{N}^*$, $i=1\dots,N$ (which may have each a finite or infinite number of elements), such that $\cup_{i=1}^N I_{i} = \mathbb{N}^*$. Then, we define the $N$ projection operators $P_i= \sum_{j\in I_{i}} P({\bf n}_j)$. According to the L\"uders-von Neumann projection formula, the possibly degenerate outcomes that are associated with them correspond to the outcome-states: 
\begin{eqnarray}
D({\bf s}_i)={P_iD({\bf r})P_i\over {\rm Tr}\, P_i D({\bf r})P_i},\quad i=1\dots,N.
\label{Luder}
\end{eqnarray}
If we introduce the vector-state notation $|\phi_i\rangle = P_i |\psi\rangle /\| P_i |\psi\rangle\|$, with $D({\bf r})=|\psi\rangle\langle\psi |$ and $D({\bf s}_i)=|\phi_i\rangle\langle\phi_i |$, we can also write the pre-measurement vector-state $|\psi\rangle$ as the superposition:
\begin{eqnarray}
|\psi\rangle = \sum_{i=1}^N \| P_i |\psi\rangle\|\, |\phi_i\rangle,
\end{eqnarray}
as is clear that $\sum_{i=1}^N P_i = \mathbb{I}$. Since ${\rm Tr}\,D({\bf s}_i)D({\bf s}_j)=\langle \phi_i|\phi_j\rangle = \delta_{ij}$, $i,j\in\{1,\dots,N\}$, we have ${\bf s}_i\cdot {\bf s}_j = \delta_{ij}$, $i,j=1,\dots,N$. This means that the $N$ (infinite-dimensional) unit vectors ${\bf s}_i$ define a standard $(N-1)$-dimensional sub-simplex of $\Delta_\infty$: 
\begin{equation}
\tilde{\triangle}_{N-1}({\bf s}_1,\dots,{\bf s}_N)=\{{\bf t}\in\real^\infty | {\bf t}=\sum_{i=1}^N t_i\, {\bf s}_i,\, \sum_{i=1}^N t_i = 1, 0\leq t_i\leq 1\}.
\label{simplex-effective}
\end{equation}
Clearly, being $\tilde{\triangle}_{N-1}({\bf s}_1,\dots,{\bf s}_N)$ finite dimensional, we have $\mu[\tilde{\triangle}_{N-1}({\bf s}_1,\dots,{\bf s}_N)]\neq 0$, so that the EBR can be consistently applied to it. 

More precisely, writing ${\bf r}= {\bf r}^\perp + {\bf r}^\parallel$, with ${\bf r}^\perp$ the component of ${\bf r}$ perpendicular to $\tilde{\triangle}_{N-1}({\bf s}_1,\dots,{\bf s}_N)$, i.e. ${\bf r}^\perp\cdot {\bf s}_i =0$, for all $i=1,\dots,N$, we have ${\bf r}^\parallel =\sum_{i=1}^N r_i^\parallel \,{\bf s}_i$, so that:
\begin{eqnarray}
{\cal P}[D({\bf r})\to D({\bf s}_i)] = {\bf r}^\parallel\cdot {\bf s}_i = r_i^\parallel,
\end{eqnarray}
and from the general properties of a simplex, we also have \cite{asdb2014}: 
\begin{eqnarray}
r_i^\parallel = {\mu (A_i) \over \mu[\tilde{\triangle}_{N-1}({\bf s}_1,\dots,{\bf s}_N)]},
\end{eqnarray}
where $A_i$ denotes the convex closure of $\{{\bf s}_1, \dots, {\bf s}_{i-1}, {\bf r}^\parallel, {\bf s}_{i+1}, \dots, {\bf s}_N\}$. In other words, for as long as the number of outcomes remains finite, even though the quantum entity is infinite-dimensional we can still describe the outcome probabilities as a condition of lack of knowledge about the measurement-interactions that are actualized at each run of the measurement.

\section{Continuous spectrum}

In the previous section, starting from an observable having a pure point spectrum, we have shown that the degenerate observables that can be built from it admit a hidden-measurement description for the transition probabilities, if the number of degenerate outcomes is finite. This seems to exclude observables also having some continuous spectrum. To show that this is not the case, in this section we present an alternative derivation, using a representation where the generators $\Lambda_i$ are constructed using the outcome states. 

More precisely, we consider a quantum entity with a possibly infinite-dimensional Hilbert space ${\cal H}$, and $N$ mutually orthogonal projection operators $P_i$, $i=1,\dots,N$, such that $\sum_i^N P_i =\mathbb{I}$. For example, if ${\cal H}=L^2(\real)$, and we consider the position observable $Q=\int_{-\infty}^\infty dx\, x\, |x\rangle\langle x|$, they could be given by the integrals: $P_i=\int_{I_i} dx\, |x\rangle\langle x|$, where the $I_i$ are disjoint intervals covering the entire real line, i.e., $\real =\cup_{i=1}^N I_i$, so that $\sum_i^N P_i =\sum_i^N \int_{I_i} dx\, |x\rangle\langle x| = \int_{-\infty}^\infty dx\, |x\rangle\langle x| = \mathbb{I}$.

If the $D({\bf s}_i)=|\phi_i\rangle\langle\phi_i |$ are the outcomes defined in (\ref{Luder}), we can use the $N$ orthonormal vector-states $|\phi_i\rangle$ to construct the first $N^2-1$ generators of $SU(\infty)$ \cite{Hioe1981}: $\{\Lambda_i\}_{i=1}^{N^2-1}=\{U_{jk},V_{jk},W_{l}\}$, with:
\begin{eqnarray}
\label{rgeneratorsN}
&&U_{jk}=|\phi_j\rangle\langle \phi_k| + |\phi_k\rangle\langle \phi_j|, \quad V_{jk}=-i(|\phi_j\rangle\langle \phi_k| - |\phi_k\rangle\langle \phi_j|),\nonumber \\
&&W_l =\sqrt{2\over l(l+1)}\left(\sum_{j=1}^l |\phi_j\rangle\langle \phi_j|-l|\phi_{l+1}\rangle\langle \phi_{l+1}|\right),\label{generatorsW}\label{W}\nonumber \\
&&1\leq j < k\leq N, \quad 1\leq l\leq N-1.
\label{generators}
\end{eqnarray}
We also define the operator $\mathbb{I}_{N}=\sum_{i=1}^N |\phi_i\rangle\langle \phi_i|$, acting as an indentity operator in the $N$-dimensional subspace ${\rm Span}\,\{|\phi_1\rangle,\dots, |\phi_N\rangle\}$. Since $|\psi\rangle\in {\rm Span}\,\{|\phi_1\rangle,\dots, |\phi_N\rangle\}$, the associated projection operator $|\psi\rangle\langle\psi|$ can be expanded on the basis $\{\mathbb{I}_{N},\Lambda_1,\dots, \Lambda_{N^2-1}\}$, and we can write:
\begin{equation}
|\psi\rangle\langle\psi|=D({\bf r})={1\over N}(\mathbb{I}_N + c_N \,{\bf r}\cdot \mbox{\boldmath$\Lambda$}) = {1\over N}\left(\mathbb{I}_N + c_N\sum_{i=1}^{N^2-1} r_i\, \Lambda_i\right).
\label{N-d-effective}
\end{equation}
Note that despite the similarity with (\ref{exp}), in (\ref{N-d-effective}) the operator-state $D({\bf r})$ is not finite-dimensional (all operators in (\ref{N-d-effective}) act in an infinite-dimensional Hilbert space ${\cal H}$). 

Considering for instance the $N=2$ case, we have the following three Pauli generators: 
\begin{equation}
\Lambda_1 = |\phi_1\rangle\langle \phi_2| + |\phi_2\rangle\langle \phi_1|,\ \Lambda_2 = -i(|\phi_1\rangle\langle \phi_2| - |\phi_2\rangle\langle \phi_1|), \ \Lambda_3=|\phi_1\rangle\langle \phi_1| - |\phi_2\rangle\langle \phi_2|,
\label{generalizedpauli}
\end{equation}
and the indentity operator $\mathbb{I}_{2}=|\phi_1\rangle\langle \phi_1| + |\phi_2\rangle\langle \phi_2|$, so that we can write:
\begin{equation}
|\psi\rangle\langle\psi|=D({\bf r})={1\over 2}(\mathbb{I}_2 + {\bf r}\cdot \mbox{\boldmath$\Lambda$}) = {1\over 2}\left(\mathbb{I}_2 + \sum_{i=1}^{3} r_i\, \Lambda_i\right).
\label{2-d-effective}
\end{equation}
For the two outcome-states we have: 
\begin{equation}
|\phi_1\rangle\langle\phi_1|=D({\bf n}_1)={1\over 2}(\mathbb{I}_2 + \Lambda_3),\quad |\phi_2\rangle\langle\phi_2|=D({\bf n}_2)={1\over 2}(\mathbb{I}_2 - \Lambda_3),
\end{equation}
which means that ${\bf n}_1=(0,0,1)$ and ${\bf n}_2=(0,0,-1)$. Note that the representation is that of a (non-standard) simplex $\Delta_1$ of measure $\mu(\Delta_1)=2$, as is clear that the two vertex vectors ${\bf n}_1$ and ${\bf n}_2$, are not orthogonal, but opposite: ${\bf n}_1=-{\bf n}_2$. 

Let us also consider, for sake of clarity, the $N=3$ case. We have then the eight Gell-Mann operators: 
\begin{eqnarray}
&\Lambda_1 = |\phi_1\rangle\langle \phi_2| + |\phi_2\rangle\langle \phi_1|,\ \Lambda_2 = -i(|\phi_1\rangle\langle \phi_2| - |\phi_2\rangle\langle \phi_1|), \ \Lambda_3=|\phi_1\rangle\langle \phi_1| - |\phi_2\rangle\langle \phi_2|\nonumber\\
&\Lambda_4 = |\phi_1\rangle\langle \phi_3| + |\phi_3\rangle\langle \phi_1|,\ \Lambda_5 = -i(|\phi_1\rangle\langle \phi_3| - |\phi_3\rangle\langle \phi_1|), \ \Lambda_6=|\phi_2\rangle\langle \phi_3| - |\phi_3\rangle\langle \phi_2|\nonumber\\
&\Lambda_7 = -i(|\phi_2\rangle\langle \phi_3| - |\phi_3\rangle\langle \phi_2|),\ \Lambda_8={1\over\sqrt{3}}(|\phi_1\rangle\langle \phi_1| + |\phi_2\rangle\langle \phi_2| -2 |\phi_3\rangle\langle \phi_3|),
\label{generalizedgellmann}
\end{eqnarray}
and the indentity operator $\mathbb{I}_{3}=|\phi_1\rangle\langle \phi_1| + |\phi_2\rangle\langle \phi_2| + |\phi_3\rangle\langle \phi_3|$, so that we can write: 
\begin{equation}
|\psi\rangle\langle\psi|=D({\bf r})={1\over 3}(\mathbb{I}_3 + \sqrt{3}\,{\bf r}\cdot \mbox{\boldmath$\Lambda$})={1\over 3}\left(\mathbb{I}_2 +\sqrt{3} \sum_{i=1}^{8} r_i\, \Lambda_i\right).
\label{3-d-effective}
\end{equation}
For the three outcome-states we have: 
\begin{eqnarray}
|\phi_1\rangle\langle\phi_1|&=& D({\bf n}_1)={1\over 3}[\mathbb{I}_3 + \sqrt{3}\, ({\sqrt{3}\over 2}\Lambda_3+{1\over 2}\Lambda_8)],\nonumber \\
|\phi_2\rangle\langle\phi_2|&=&D({\bf n}_2)={1\over 3}[\mathbb{I}_3 + \sqrt{3}\, (-{\sqrt{3}\over 2}\Lambda_3+{1\over 2}\Lambda_8)],\nonumber \\
|\phi_3\rangle\langle\phi_3|&=&D({\bf n}_3)={1\over 3}[\mathbb{I}_3 + \sqrt{3}\, (-1) \Lambda_8],
\end{eqnarray}
and the associated $8$-dimensional unit vectors are: 
\begin{eqnarray}
&&{\bf n}_1=(0,0,{\sqrt{3}\over 2},0,0,0,0,{1\over 2}),\quad 
{\bf n}_2=(0,0,-{\sqrt{3}\over 2},0,0,0,0,{1\over 2}),\nonumber \\
&&{\bf n}_3=(0,0,0,0,0,0,0,-1),
\end{eqnarray}
which clearly form an equilateral triangle, that is, a 2-simplex $\Delta_2$. 

We thus see that an EBR of the measurement context is still possible, if the latter only involves a finite number of outcomes, which can also correspond to operators projecting onto some continuous spectrum of the observable under consideration. For this, the outcome-states have to be used to construct the first $N^2-1$ generators $\Lambda_i$, which means that we have now to renounce using a same representation to describe different measurements, unless they would all produce outcomes belonging to the same subspace ${\rm Span}\,\{|\phi_1\rangle,\dots, |\phi_N\rangle\}$. 

Let us illustrate this last observation in the simple $N=2$ case. We consider a measurement whose outcome-states are $|\phi'_1\rangle$ and $|\phi'_2\rangle$, which we assume also form a basis of ${\rm Span}\,\{|\phi_1\rangle,|\phi_2\rangle\}$. We can then generally write: $|\phi'_1\rangle= u_{11}|\phi_1\rangle + u_{12}|\phi_2\rangle$, and $|\phi'_2\rangle= u_{21}|\phi_1\rangle + u_{22}|\phi_2\rangle$. The condition $\langle \phi'_1|\phi'_1\rangle =1$ implies: $|u_{11}|^2+|u_{12}|^2=1$, and the condition $\langle \phi'_2|\phi'_2\rangle =1$ implies: $|u_{21}|^2+|u_{22}|^2=1$. Also, condition $\langle \phi'_1|\phi'_2\rangle =0$ implies: $(u^*_{11}\langle\phi_1| + u^*_{12}\langle\phi_2|)(u_{21}|\phi_1\rangle + u_{22}|\phi_2\rangle)=0$, i.e., $u^*_{11}u_{21}+u^*_{12}u_{22}=0$. Thus, the $2\times 2$ matrix $U$, with elements $[U]_{ij}=u_{ij}$, obeys: 
\begin{equation}
UU^\dagger =\left[ \begin{array}{cc}
u_{11} & u_{12} \\
u_{21} & u_{22} \end{array} \right] \left[ \begin{array}{cc}
u^*_{11} & u^*_{21} \\
u^*_{12} & u^*_{22} \end{array} \right] =\left[ \begin{array}{cc}
1 & 0 \\
0 & 1 \end{array} \right]. 
\end{equation}
We have: 
\begin{eqnarray}
|\phi'_1\rangle\langle\phi'_1|&=&D({\bf n}'_1)=(u_{11}|\phi_1\rangle + u_{12}|\phi_2\rangle)(u^*_{11}\langle\phi_1| + u^*_{12}\langle\phi_2|)\\
&=& |u_{11}|^2 D({\bf n}_1) + |u_{12}|^2 D({\bf n}_2) + u_{11}u^*_{12}|\phi_1\rangle\langle\phi_2| + u_{12}u^*_{11}|\phi_2\rangle\langle\phi_1|\nonumber\\
&=&{1\over 2}[\mathbb{I}_2 + (|u_{11}|^2-|u_{12}|^2)\Lambda_3 +2u_{11}u^*_{12}|\phi_1\rangle\langle\phi_2| + 2u_{12}u^*_{11}|\phi_2\rangle\langle\phi_1|]\nonumber\\
&=&{1\over 2}[\mathbb{I}_2 + (|u_{11}|^2-|u_{12}|^2)\Lambda_3 +(u_{11}u^*_{12} +u_{12}u^*_{11})\Lambda_1 -i(u_{11}u^*_{12} -u_{12}u^*_{11})\Lambda_2 ].\nonumber
\end{eqnarray}
In other words, the components of ${\bf n}'_1$ are: 
\begin{equation}
{\bf n}'_1=(2\Re (u_{11}u^*_{12}), 2\Im (u_{11}u^*_{12}), |u_{11}|^2-|u_{12}|^2),
\end{equation}
and we can check that ${\bf n}'_1\cdot {\bf n}'_1=4|u_{11}|^2|u_{12}|^2 + (|u_{11}|^2-|u_{12}|^2)^2= (|u_{11}|^2+|u_{12}|^2)^2=1$. Of course, a similar calculation can be done to determine the coordinates of ${\bf n}'_2$, associated with $|\phi'_2\rangle\langle\phi'_2|$. So, it is possible to describe, within the same 3-dimensional effective Bloch sphere, all two-outcome measurements with outcome-states $|\phi'_1\rangle$ and $|\phi'_2\rangle$ belonging to ${\rm Span}\,\{|\phi_1\rangle,|\phi_2\rangle\}$, i.e., of the form:
\begin{equation}
\left[ \begin{array}{c}
|\phi'_1\rangle \\
|\phi'_2\rangle \end{array} \right] =
\left[ \begin{array}{cc}
u_{11} & u_{12} \\
u_{21} & u_{22} \end{array} \right] \left[ \begin{array}{c}
|\phi_1\rangle \\
|\phi_2\rangle \end{array} \right].
\end{equation}

\section{Conclusion}
\label{Conclusion}

In this article, we emphasized that when we take the infinite-dimensional limit of the EBR we face the problem that the Lebesgue measures of the simplexes tend to zero, so preventing the direct use of the infinite EBR to express the outcome probabilities as relative measures of the simplexes' sub-regions. However, we have also shown that the problem can be overcome by observing that measurements are operations that in practice always present a finite number of outcome-states (possibly associated with an infinite dimension of degeneracy), so that their representation only requires finite-dimensional simplexes. 

In other words, we have proposed to distinguish the dimension of a quantum entity \emph{per se}, expressing its `intrinsic potentiality', which can either be finite or infinite, from the dimension of a measurement (the number of outcomes that are available in our spatiotemporal theater, in a given experimental situation), which determines the `effective potentiality' that can be manifested by a quantum entity, when submitted to the former. In practice, the dimension of an actual interrogative context is always finite, as the number of macroscopic entities that can play the role of detectors is finite and their resolving powers are limited. In that respect, one could even go as far as saying, albeit only speculatively, that the measurement-interactions responsible for the transitions to the different possible outcome-states in fact supervene and produce their effects only when the (possibly infinite) potentiality level associated with the quantum entity gets constrained by a finite number of possible outcomes, during the practical execution of a measurement.

To conclude, let us offer an analogy taken from the domain of human cognition. Consider a person submitted to an interrogative context, forced to choose one among a finite number of distinct answers. For this, the person's mind has to immerse into the semantic context created by the question and the available answers, gradually building up a tension with each one of them; a tension that in the end will have to be released, thus producing an outcome.\footnote{In quantum cognition, a more general version of the EBR, not limited to the Hilbert geometry for the state space, called the general tension-reduction (GTR) model, has been proposed to model human decision processes~\cite{asdb2015c}.} However, if the number of possibilities to be taken simultaneously into account in providing an answer increases, it will become more and more difficult for the person's mind to maintain a sufficient cognitive interaction with each one of them. In other words, in the limit where the number of possible answers tends to infinity, the cognitive interactions associated with each one of them will either tend to zero, and then the process of actualization of an answer cannot take place (note that measurements might as well not produce an outcome), or the person's mind will start focusing on a finite subset of possibilities, with respect to which a tension-reduction process, yielding an outcome, can again take place.

\section*{Appendix: The standard simplex representation}
\label{Appendix}

To better understand what happens when one takes the infinite limit, one can adopt from the beginning a representation where the scalar product between the different vertex vectors is independent of the dimension $N$, and such that ${\bf 0}\notin \triangle_{N-1}$. Indeed, in the representation used in \cite{asdb2014}, which naturally emerges from the Hilbert geometry, the scalar product between the different vertex vectors is given by 
(\ref{scalar}), thus it depends on the dimension $N$. To eliminate this dependency, one can first introduce $N$ mutually orthogonal vectors ${\bf m}_i$, $i=1,\dots,N$, of length $\sqrt{N}$, such that (see Fig.~\ref{vectors}, for the $N=2$ case): 
\begin{figure}[!ht]
\centering
\includegraphics[scale =.35]{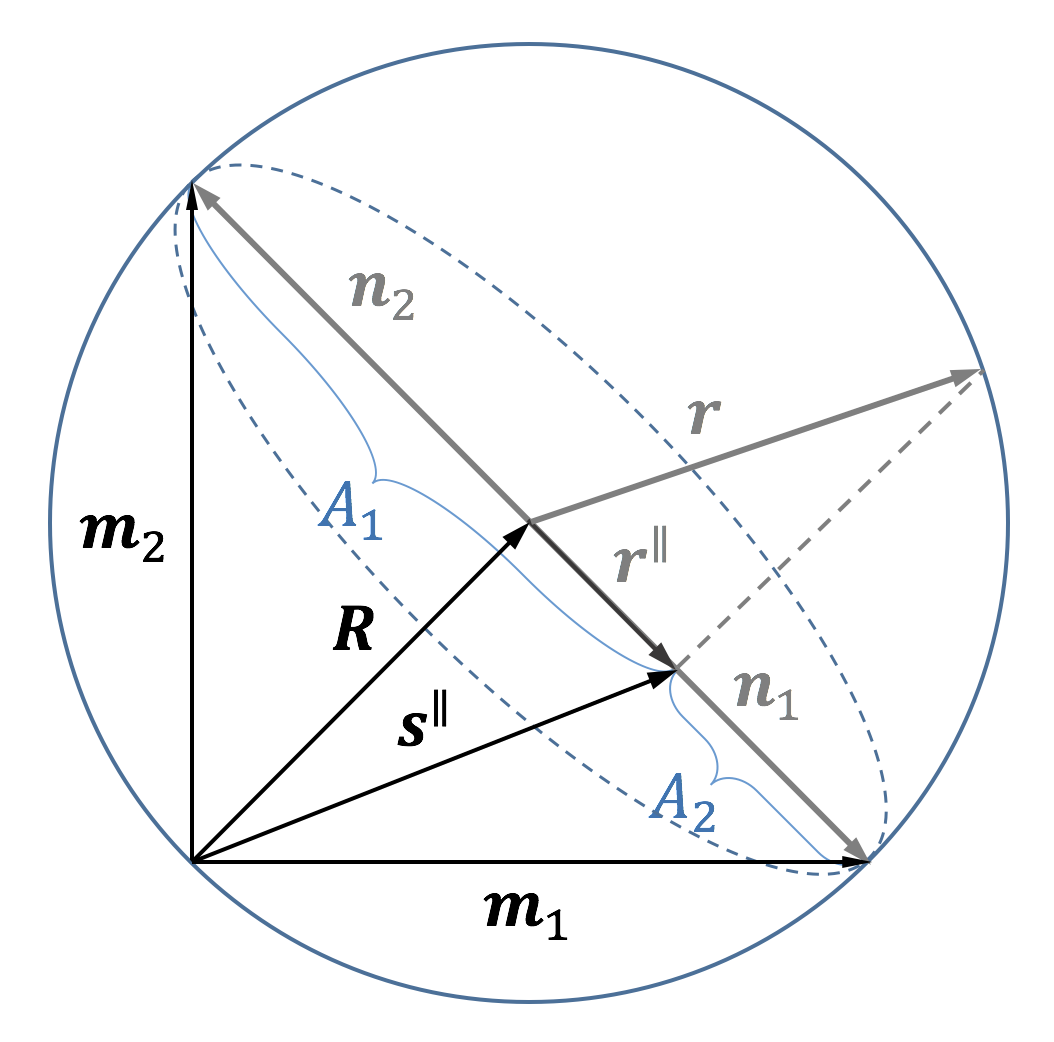}
\caption{For the $N=2$ case, the following vectors are represented in the figure: the unit vector ${\bf r}$, describing the initial state of the entity, the orthogonally projected vector ${\bf r}^\parallel$, describing the on-simplex state, the two unit vectors ${\bf n}_1$ and ${\bf n}_2$, describing the two vertices of the one-dimensional measurement simplex $\triangle_{1}$ (corresponding to a diameter of the 3-dimensional Bloch sphere), the two orthogonal vectors ${\bf m}_1$ and ${\bf m}_2$, of length $\sqrt{2}$, the unit vector ${\bf R}={1\over 2}({\bf m}_1 + {\bf m}_2)$, corresponding to the center of the simplex and of the Bloch sphere, and the translated vector ${\bf s}^{\parallel}={\bf r}^\parallel + {\bf R}$. 
\label{vectors}}
\end{figure} 
\begin{eqnarray}
&{\bf n}_i= \sqrt{1\over N-1}({\bf m}_i -{\bf R}),\quad {\bf R}={1\over N}\sum_{i=1}^N {\bf m}_i,\quad {\bf m}_i\cdot{\bf m}_j=N\delta_{ij}.
\label{relation}
\end{eqnarray}
We then have $\|{\bf R}\|^2 =1$ and ${\bf m}_i \cdot {\bf R}=1$, for all $i$, and one can check that, in accordance with (\ref{scalar}), ${\bf n}_i\cdot{\bf n}_j = {1\over N-1}({\bf m}_i \cdot {\bf m}_j -{\bf m}_i \cdot{\bf R} -{\bf m}_j \cdot{\bf R} +\|{\bf R}\|^2)= {1\over N-1}(N\delta_{ij} -1 -1+1)= \delta_{ij}{N\over N-1} -{1\over N-1}$. Introducing also the vector ${\bf s}^{\parallel}$, defined by: ${\bf r}^\parallel= \sqrt{1\over N-1}({\bf s}^{\parallel} -{\bf R})$, 
one finds for the transition probability (\ref{transitiongeneralNxN}): 
\begin{eqnarray}
{\cal P}[D_N({\bf r})\to P_N({\bf n}_i)] &=& {1\over N} [1+ ({\bf s}^{\parallel} -{\bf R})\cdot ({\bf m}_i -{\bf R})]\nonumber \\
&=& {1\over N} [1+ ({\bf s}^{\parallel}\cdot {\bf m}_i - {\bf s}^{\parallel}\cdot {\bf R} - {\bf m}_i\cdot {\bf R}+ \|{\bf R}\|^2)]\nonumber \\
&=& {1\over N} [1+ ({\bf s}^{\parallel}\cdot {\bf m}_i - 1 - 1+1)]= {1\over N}\, {\bf s}^{\parallel}\cdot {\bf m}_i,
\label{transitiongeneralNxN2}
\end{eqnarray}
where we have used the fact that ${\bf R}\cdot {\bf n}_i =0$, for all $i=1,\dots,N$, so that ${\bf R}\cdot {\bf r}^{\parallel} =0$, and consequently ${\bf s}^{\parallel}\cdot {\bf R}= (\sqrt{N-1}\, {\bf r}^{\parallel} + {\bf R})\cdot {\bf R}=1$. Then, one can introduce the unit vectors $\tilde{\bf m}_i=\sqrt{1\over N}\,{\bf m}_i$, associated with the standard simplex:
\begin{equation}
\tilde\triangle_{N-1}=\{{\bf t}\in\real^N | {\bf t}=\sum_{i=1}^{N} t_i\, \tilde{\bf m}_i,\, \sum_{i=1}^N t_i = 1, 0\leq t_i\leq 1\},
\label{simplex2}
\end{equation}
to which belongs the renormalized vectors $\tilde{\bf R}=\sqrt{1\over N}\,{\bf R}={1\over N}\sum_{i=1}^N \tilde{\bf m}_i$ and ${\tilde{\bf s}}^{\parallel}=\sqrt{1\over N}\,{\bf s}^{\parallel}$, so that (\ref{transitiongeneralNxN2}) simply becomes: 
\begin{equation}
{\cal P}[D_N({\bf r})\to P_N({\bf n}_i)] = \tilde{\bf s}^{\parallel}\cdot \tilde{\bf m}_i,
\label{transitiongeneralNxN2-bis}
\end{equation}
i.e., does not depend anymore explicitly on the dimension $N$, so that the $N\to\infty$ limit becomes trivial. Clearly, $\triangle_{\infty}$ is the limit of both $\tilde{\triangle}_{N-1}$ and ${\triangle}_{N-1}$, considering that $\tilde{\bf R}\to {\bf 0}$, as $N\to\infty$, so that in this limit one also has: ${\bf n}_i=\sqrt{N\over N-1}(\tilde{\bf m}_i -\tilde{\bf R})\to \tilde{\bf m}_i$ and ${\bf r}^\parallel=\sqrt{N\over N-1}(\tilde{\bf s}^{\parallel} -\tilde{\bf R})\to\tilde{\bf s}^{\parallel}$.

\end{document}